*Mechanism of silica-lysozyme composite formation unravelled by in situ fast SAXS*

Tomasz M. Stawski*[1,2], Daniela van den Heuvel[2,3], Rogier Besselink[1], Dominique J. Tobler[4], Liane G. Benning[1,2,5]

[1]German Research Centre for Geosciences, GFZ, Interface Geochemistry, Potsdam, Germany;

[2]School of Earth and Environment, University of Leeds, Leeds, UK;

[3]Rock-Water Interaction Group, Institute of Geological Sciences, University of Bern, Bern, Switzerland;

[4]Nano-Science Centre, Department of Chemistry, University of Copenhagen, Copenhagen, Denmark;

[5]Geochemistry Group, Department of Earth Sciences, Free University of Berlin, Berlin, Germany.

**\*** corresponding author: stawski@gfz-potsdam.de

**Abstract**

A quantitative understanding of aggregation mechanisms leading to the formation of inorganic nanoparticles (NPs) and protein composites in aqueous media is of paramount interest for colloid chemistry. In particular, the interactions between silica ($SiO_2$) NPs and lysozyme (LZM) have attracted attention, because LZM is well-known to adsorb strongly to silica NPs, while at the same time preserving its enzymatic activity. The inherent nature of the aggregation processes leading to NP-LZM composites involves structural changes at length-scales from few to hundreds of nanometres but also time scales << 1 second. To unravel these we used *in situ* synchrotron-based small-angle X-ray scattering (SAXS) and followed the subtle interparticle interactions in solution at a time resolution of 50 ms/frame (20 fps). We show that if the size of silica NPs (~5 nm diameter) is matched by the dimensions of LZM, the evolving scattering patterns contain a unique structure factor contribution originating from the presence of LZM. We developed a scattering model and applied it to analyse this structure function, which allowed us to extract structural information on the deformation of lysozyme molecules during aggregation, as well as to derive the mechanisms of composite formation.

**Introduction**

A mechanistic understanding of aggregation in aqueous media leading to the formation of inorganic nanoparticles and protein composites is of paramount interest for colloid chemistry, Earth sciences (e.g. biomineralisation) or the design of biomedical devices and sensors[1–4]. In particular, amorphous silica ($SiO_2$) nanoparticles (NPs) and lysozyme (LZM) composites have attracted attention because silica NPs readily form in many Earth surface environments (e.g., oceans, hot springs) where biological activity dominates (e.g., diatom formation) but they are also key components in numerous technological applications from electronics to paint production. In turn, lysozyme adsorbs strongly to silica NPs[5–7], while at the same time preserving its enzymatic activity most notably antibacterial properties[8]. Over a broad range of pH values (2-~10), the surface of silica NPs is negatively charged, whereas the LZM molecule is positively charged[9,10]. This way LZM can act as a bridge between silica NPs, leading to aggregation and flocculation and thus to large silica NP-LZM composites. In the $SiO_2$-LZM model system, a number of studies investigating the relationship between silica NP sizes, and adsorption modes of lysozyme revealed a correlation between composite properties and the protein's folding/structure, its enzymatic activity and the exact protein localization with respect to the silica NPs inside flocculated composites[5,6,8,10–13]. The inherent nature of the aggregation processes leading to silica-lysozyme composites, involves structural changes at length-scales from a few to hundreds of nanometres, which makes this system well-suited to be characterized by scattering methods. In particular, recent advances[14,15] in detector technology for synchrotron-based *in situ* and time-resolved small-angle X-ray scattering (SAXS), now allows to follow all steps in the formation of $SiO_2$-LZM composites from the individual components. Nevertheless, although such scattering data will reflect the *in situ* state of a system during measurement, the quantitative information related to any changes in structural properties of the particles/species of interest can only be accessed by developing, testing and validating relevant models and bespoke data analysis methods.

In this study we show step-by-step how a scattering model was developed, verified and applied to time-resolved synchrotron-based SAXS data in which we followed *in situ* the lysozyme induced aggregation of silica NPs (~5 nm in diameter) at a time resolution of 50 ms. It is important to note that at typical NPs sizes and concentrations[6,13] and protein concentrations[16] usually employed to make such $SiO_2$-LZM composites, the relative X-ray scattering contrast/intensity of lysozyme, in the presence of silica, is insufficient to determine the protein contribution to the overall scattering pattern (ratio of ~1:100 between LZM and NPs), and only the NP component can directly be followed. However, through this study we demonstrate that if the size of the silica NPs is matched with the dimensions of lysozyme (ellipsoidal molecule[17] 3 x 3 x 4.5 $nm^3$), the evolving scattering patterns contain a unique structure factor contribution originating from the presence of lysozyme, and this way this important contribution can be assessed. The analysis of this structure function through the derived model then allowed us to extract detailed structural information on the deformation of the LZM molecules upon aggregation, and to determine the mechanisms of $SiO_2$-LZM composites composite formation.

**Experimental**

*Synthesis of amorphous silica lysozyme-composites:*

Sodium metasilicate ($Na_2SiO_3 \cdot 5H_2O$, technical grade), hen egg white lysozyme (crystalline, powdered) and HCl (37%, analytical grade) were purchased from Sigma Aldrich. Separate stock solutions of dissolved silica ($SiO_2$ =1000 ppm, pH=12.5) and lysozyme (5 wt.%, pH=3.5) were prepared by dissolving the required amount of sodium metasilicate or lysozyme in ultrapure deionised water (18.2 MΩ·cm). Silica NPs were prepared in a 500 mL plastic reactor by neutralizing the silica stock solution through titration with HCl until a pH of 7.5 was reached. This neutralized solution was left to polymerize and age for 16 hours. Silica-protein composites were obtained by mixing the silica NP solution with a pre-measured amount of the lysozyme stock solution under rapid stirring (500 rpm) to yield a $SiO_2$ NPs solution with 1000 ppm lysozyme (final pH 6.9, salinity 20 mM).

*Scattering experiments:*

The formation process and the development of the structure of the silica-LZM composites was studied *in situ* and in a time-resolved manner by synchrotron-based small angle X-ray scattering (SAXS) at the BioSAXS beamline[18] P12 of the EMBL at PETRA III (DESY, Germany) using a monochromatic X-ray beam at 10 keV. Two-dimensional scattered intensities were collected at small-angles with a Dectris Pilatus 2M (2D large area pixel-array detector) using an acquisition time of 50 ms per frame. Transmission was measured by means of a photodiode installed in the beam-stop of the SAXS detector. A sample-to-detector distance of ~3 m allowed for a usable q-range of ~ $0.04 < q < 4.5$ $nm^{-1}$. The scattering-range at small-angles was calibrated against silver behenate, and the intensity was calibrated to absolute units against water. For the *in* situ experiment, first, the starting silica NP solution was continuously circulated between the reactor (where the suspension was stirred at 500 rpm) and the flow-through cell with embedded quartz capillary (ID 1.7 mm, wall thickness 50 μm; aligned with the X-ray beam) using a peristaltic pump (Gilson MiniPuls 3, flow ~500 mL/min; tubing: ID 2 mm, total length 2 m; reactor-to-cell-distance: 0.7 m of tubing). Once a SAXS baseline for the silica NP solution was recorded, the pre-measured amount of lysozyme stock solution was pumped into the reactor at a fast rate. This injection was done remotely from the operator hutch via a 10 m long PTFE tube (ID 4 mm) that was routed into the reactor located in the experimental hutch. The tube was filled in such a way that the LZM solution was located in the last ~40 cm of the tube on the reactor side. The other end of the tube in the operators' room was equipped with a 50 mL syringe filled with air. Thus the experiment started with recording of 24 s (480 x 50 ms) SAXS patterns of the silica NP solution circulating through the capillary, prior to the fast injection of the entire content of the tube containing the lysozyme with a single rapid push of the syringe plunger that lasted ~200 - 400 ms. This fast injection rate in combination with the fast stirring in the reactor (500 rpm), pumping (500 mL/min) and fast data acquisition (50 ms / SAXS pattern) provided best-possible conditions for the characterisation of all the steps leading to the formation of the silica – LZM composites. The used experimental set-up introduced a unavoidable dead-time of ~500 ms between the injection moment and the first actual measurement of the mixed solution i.e., the time required for the mixed solution to reach the capillary where the SAXS pattern was recorded. In order to be able to analyse and model the silica–LZM composite scattering patterns we also acquired a series of backgrounds and reference samples including an empty capillary and a capillary filled with water, silica stock solution, LZM solutions at different concentrations. The initial SAXS data processing and reduction included a series of automatic post-data collection steps including masking of undesired pixels, normalizations and correction for transmission, instrumental background subtraction and data integration of the collected 2D data to 1D. Further data processing and water background subtraction, model fitting, validation and analysis, were performed through a bespoke scripts developed in GNU Octave[19,20]. The script we developed as well as all the documentation and the selected scattering curves are available at: https://github.com/tomaszstawski/SilicaLysozymeSAXS. In a first instance for the model, we obtained the size distribution of the initial silica NPs from a Monte Carlo fitting implemented[21,22] in MCSAS under the assumption that the silica NPs particles were spherical[23].

*Characterization of dry samples:*

To cross-correlate the *in situ* SAXS data, the silica-lysozyme suspensions were dried in an oven at 40 °C for ~ 48 hours. The resulting powders were washed 5 times with MilliQ water to remove excess lysozyme and salts followed by a 2$^{nd}$ drying step at 40 °C. The amount of lysozyme associated with the composites was quantified by determining the total carbon content in solids by mass spectrometry (DELTAplusXL ThermoFisher) with a Carlo-Erba NC2500. From these analyses the lysozyme content was calculated using the molecular formula $C_{613}H_{959}N_{193}O_{185}S_{10}$ and molecular weight of 14313 g/mol for lysozyme[24] (ProtParam based on UniProtKB entry P00698).

**Results and Discussion**

*Evolution of SAXS patterns and derived aggregation stages*

Upon mixing of the silica NPs and the LZM solution we observed very fast flocculation, which indicated the formation of the composites. In Fig. 1 we show an overview of these formation processes based on SAXS data collected at a time resolution of 50 ms and spanning ~300 s. In a contour plot of the time-resolved scattering patterns (Fig. 1A) one can distinguish 4 characteristic time periods (I-IV) and one region of interest (ROI V), which spanned through periods II to IV. Period I corresponds to the initial ~24 s of the scattering patterns of silica NPs before the injection of lysozyme. Based on this data we determined the initial form factor (size distribution) of the silica NPs prior to mixing with lysozyme (Fig. 1B). Since the scattering pattern in a Porod representation prominently flattened out at low-$q$ (i.e. $I(q) \propto q^0$) this shows that the starting silica NPs were not aggregated and well-suspended. We derived a discrete size distribution (histogram in the inset, Fig. 1B) for the NPs from the Monte Carlo fit implemented[21] in MCSAS under the *a priori* assumption that the NPs were spherical in shape[23] (physicochemical parameters of amorphous silica given in Table A1). The as-obtained histogram indicated that the size distribution was relatively narrow with a mean radius of 2.53±0.01 nm (distribution statistics given in Table A2). The total integrated volume fraction for the NPs obtained from the fit, was 0.040±0.001%, which matches very closely the expected value of 0.041% calculated for silica NP precipitated from a 1000 ppm $SiO_2$ solution at pH 7.5 and 21 deg. C (Table A1, calculated with PHREEQC[25]).

Period II (~25 – ~30 s) in Fig. 1A represents scattering patterns during and soon after the injection of the LZM solution and its mixing with the silica NPs. Period II is hence preceded by a 500-ms-dead-time period (see Experimental). Period II (Fig. 1C) can be divided into multiple steps. The first 1.25 s were primarily characterized by a rapid and significant (~11 fold) increase in intensity at low-$q$ ($q < 0.3$ nm$^{-1}$). During the following 2 s, the low-$q$ part still kept increasing (to ~15 times the initial intensity) but less rapidly, and at $q \sim 1$ nm$^{-1}$ a characteristic local maximum developed (ROI V in Fig. 1A). The intensity increase at low-$q$ originated from the formation of large aggregates constituting the composites, with sizes outside the minimum $q$-range, whereas the local maximum (the correlation peak $q \sim 1$ nm$^{-1}$) indicated the presence of interparticle correlations within those aggregates.

The intensity increase at low-$q$ associated with the aggregation continued throughout period III (between ~30 and 150 s), yet the correlation peak in ROI V did not change significantly (Fig. 1D).

Note that up to 150 s (periods I-III) the high-$q$ part of the data ($q > \sim 1.5$ nm$^{-1}$, Fig. 1C, D) did not change, indicating that the original form factor of silica NPs remained the same after the injection of lysozyme. Therefore, as a first approximation the observed electron density scattering contrast in these periods (I-III) can be interpreted to originate solely from the silica NPs and not from the combination of silica and lysozyme (Fig. A1). Hence, we could treat the system as a 2-electron-density system (silica NPs and solvent matrix). However, because our silica nanoparticles were smaller (~5 nm) compared to silica NPs in previous scattering studies (~20 nm) on silica-protein composite formation[5,6,12,13] and because the lysozyme addition dramatically changed the silica aggregation state, the contributions of the lysozyme scattering can be accounted for indirectly from the interparticle correlations observed in ROI V. This is key here, because it allows us to extract the changes in lysozyme structural properties as the composites evolve over time, without deriving circumstantial models for a 3-electron-density system (silica, lysozyme, and the solvent matrix).

In period IV (150 – 300 s, Fig. 1A and 1D), we observed a further intensity increase at low-$q$ (3 times higher at 300 s than at 150 s), which indicated a continuous increase in aggregate size from periods II and III. In this time-period IV the intensity of the ROI V (Fig. 1A) started to increase together with the silica form factor at high-$q$ ($q > \sim 1.5$ nm$^{-1}$, Fig. 1D). This suggests that as aggregation continued between 150 and 300 s, the silica NPs themselves started to grow e.g. due the coalescence of NPs or similar processes (under an assumption that the particles remained spherical in shape).

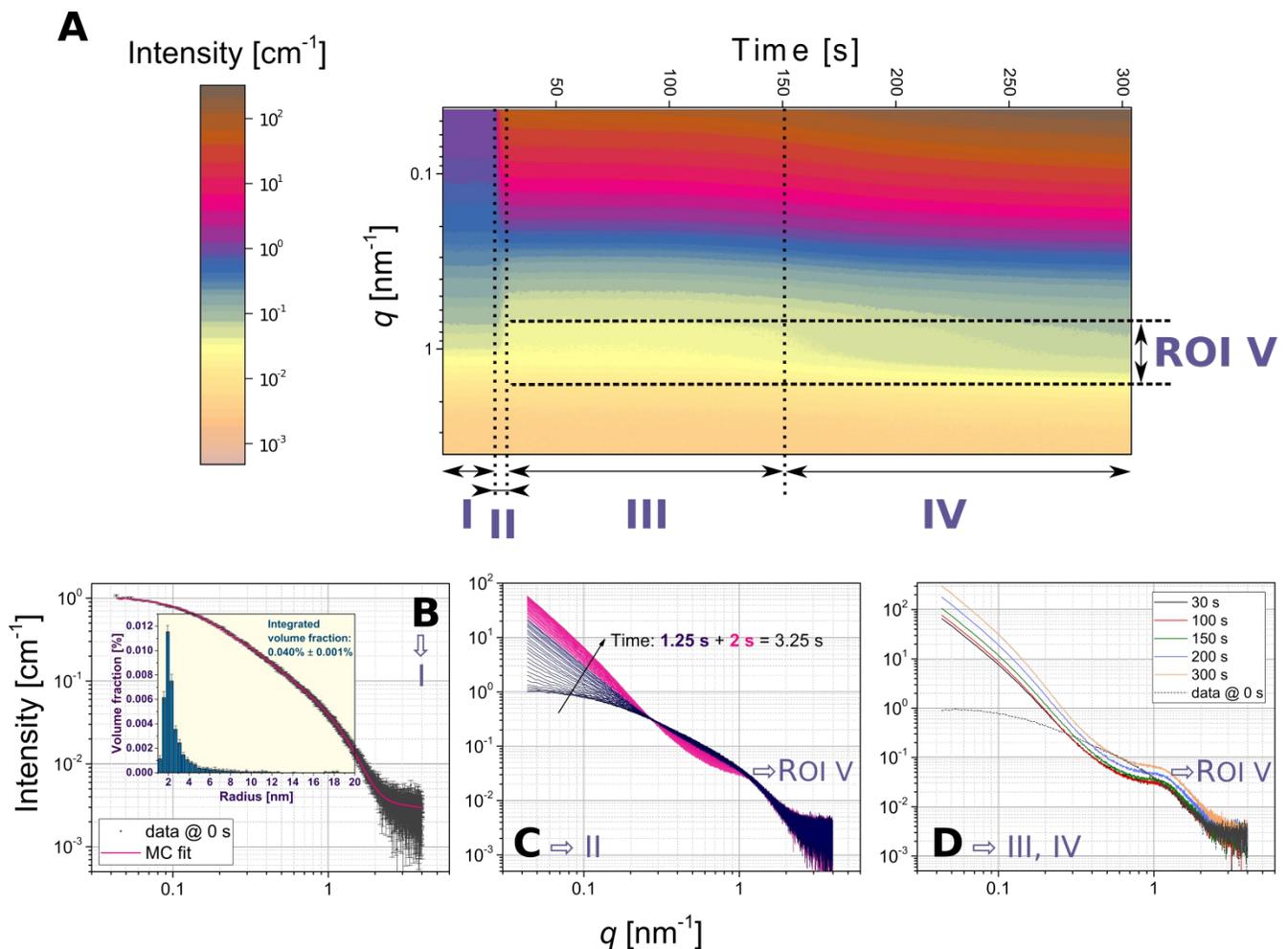

**Figure 1.** Time-resolved and *in situ* SAXS patterns documenting the formation of silica-lysozyme aggregates from an experiment where data was recorded at a rate of 50 ms/pattern (20 fps); A) contour plot depicting the scattering intensity as a function of *q* and time. The changes in the data reveal four distinct time periods; I – scattering from pure silica NPs, II – the moment of lysozyme injection, III – the growth of the aggregates/composites with the original silica NPs' form factor preserved, and IV – further growth of aggregates and a change in the original silica NPs. We further identified a *q*-range as a region of interest "ROI V" indicating a local maximum due to interparticle correlations; B) the initial silica NP form factor with a Monte Carlo (MC) fit and the derived discrete particle size distribution (inset); C) the scattering data for period II showing the time frames spanning the injection of the LZM solution between 24.25 s and 27.50 s of the experiment, with the first 1.25 s (blue) dominated by the formation of large aggregates, followed by ~ 2 s (pink) during which the local maximum related to the interparticle interactions (ROI V) clearly developed; D) selected patterns spanning periods III and IV highlighting the differences in the silica particles' form factor ($q > 1.5$ nm$^{-1}$). In C&D the data uncertainties are not shown for clarity.

*Scattering model*

In the considered silica-LZM composites, the scattering contrast originating from the lysozyme itself can be mostly disregarded (see Fig. A1). Hence, the scattering contrast of the primary silica particles, $(\Delta\rho_p)^2$ (expressed as the difference in scattering length density, SLD) is equal to the squared difference in the SLDs of silica and the surrounding water matrix (Table A1). Consequently, using such an approximation allows us to deal with a 2-electron-density-system, where the scattering intensity, $I(q)$ is a product of a scattering contrast, $(\Delta\rho_p)^2$, the silica NPs' form factor $P_p(q)$, weighted by a volume fraction of silica NPs, $\phi_p$, and their volume $V_p$; and an effective structure factor function describing the spatial arrangement of silica NPs within the aggregates $S_{eff}(q)$ (Eq. 1). We use subscript "p" to emphasize that the parameters and functions concern the primary silica NPs.

$$I(q) = (\Delta\rho_p)^2 \cdot \phi_p \cdot V_p \cdot P_p(q) \cdot S_{eff,p}(q) \qquad (1)$$

Eq. 1 is valid only for a system of ideally monodisperse particles (i.e., the distribution is a delta function), which is actually not the case for the silica NPs used here (Fig. 1B). This is an important consideration when including the interparticle interactions from the structure factor. There are several approaches to consider particles' polydispersity together with a structure factor[26], but because the fitted size distribution (histogram in Fig. 1B) is discrete with a finite number of *n* bins a Local Monodisperse Approximation[27] (LMA) is used in our models:

$$I(q) = (\Delta\rho_p)^2 \sum_{i=1}^{n} \left( \phi_{p,i}(r_i) \cdot V_p(r_i) \cdot P_p(q, r_i) \cdot S_{eff,p}(q, r_i) \right) \qquad (2)$$

where $P_p(q,r_i)$ is a form factor of a sphere of radius $r_i$.

The discrete size distribution (Fig. 1B) has a form of Eq. 3, where for each size contribution ($r_i$) the corresponding volume fractions ($\phi_i$) are known.

$$D(r_i) = \sum_{i=1}^{n} \phi_{p,i}(r_i) \qquad (3)$$

Since the partial *i*-th structure factor $S_{eff}(q,r_i)$ includes the interparticle correlations between silica NPs and lysozyme (the local maximum at $q_{max} \sim 1$ nm$^{-1}$, ROI V in Fig. 1), it is also dependent on $r_i$ of the primary silica NPs. Additionally, the structure factor expression has to account for the aggregation of the silica NPs to large objects (the low-$q$ increase), yet the size of these aggregates is in turn not necessarily dependent on the size of the primary silica NPs.

*The interparticle correlations and the local maximum*

Under the considered physicochemical conditions the inorganic silica NPs and the protein molecules are oppositely charged[10], and hence they interact through the attractive potential. This in turn leads to the formation of aggregates/composites in which NPs behave closely to adhesive hard spheres. The aforementioned interactions and the arrangement of particles in space lead to the occurrence of a broad maximum in the scattering pattern (Fig. 1, ROI V). We simulate these effects by implementing the adhesive hard sphere structure factor $S_{SHS}(q)$[28–30]. The interaction between particles at the distance $x$ is approximated by the following potential $U(x)$:

$$\frac{U(x)}{k_B T} = \begin{cases} \infty \Rightarrow 0 < x < 2R_{HS} \\ \ln\left(12\tau\Delta/(2R_{HS}+\Delta)\right) \Rightarrow 2R_{HS} < x < 2R_{HS} + \Delta \\ 0 \Rightarrow x > 2R_{HS} + \Delta \end{cases} \qquad (4)$$

Where $R_{HS}$ is a hard sphere radius of particles, $\Delta$ is a width of a potential well, and $\tau$ is stickiness parameter. The structure factor expression $S_{SHS}(q)$ is defined through the following set of equations (Eq. 5).

$$S_{\text{SHS}}(q, R_{\text{HS}}, v, \tau) = \frac{1}{1 - C(q)} \tag{5a}$$

$$\kappa = 2qR_{\text{HS}} \tag{5b}$$

$$\eta = v\left(\frac{2R_{\text{HS}} + \Delta}{2R_{\text{HS}}}\right)^3 \tag{5c}$$

$$\varepsilon = \tau + \frac{\eta}{1 - \eta} \tag{5d}$$

$$\gamma = v\frac{1 + \eta/2}{3(1 - \eta)^2} \tag{5e}$$

$$\lambda = \frac{6}{\eta}\left(\varepsilon - \sqrt{\varepsilon^2 - \gamma}\right) \tag{5f}$$

$$\mu = \lambda\eta(1 - \eta) \tag{5g}$$

$$\alpha = \frac{(1 + 2\eta - \mu)^2}{(1 - \eta)^4} \tag{5h}$$

$$\beta = -\frac{3\eta(2 + \eta)^2 - 2\mu\left(1 + 7\eta + \eta^2\right) + \mu^2(2 + \eta)}{2(1 - \eta)^4} \tag{5i}$$

$$C(q) = 2\frac{\eta\lambda}{\kappa}\sin\kappa - 2\frac{\eta^2\lambda^2}{\kappa^2}(1 - \cos\kappa) \tag{5j}$$

$$-\{\alpha\kappa^3(\sin\kappa - \kappa\cos\kappa) + \beta\kappa^2(2\kappa\sin\kappa - (\kappa^2 - 2)\cos\kappa - 2)$$

$$+\frac{\eta\alpha}{2}\left((4\kappa^3-24\kappa)\sin\kappa-(\kappa^4-12\kappa^2+24)\cos\kappa+24\right)\}\cdot 24\frac{\eta}{\kappa^6}$$

In this study, stickiness $\tau$, is calculated from Eq. 4, based on the literature data for the silica-LZM composites synthesized under similar physicochemical conditions[13]: $U(2R_{HS} < x < 2R_{HS} + \Delta)$ = -2.5 $k_BT$; $\Delta = 0.1\cdot(2R_{HS})$ under the restriction that $\Delta_{min} \geq 0.15$ nm (*i.e.,* the average H-bond length). The value of $U$ in our study may slightly differ from literature values, because for the constant pH and salinity, the surface charge of NPs increases with decreasing size[31]. However, it is unlikely that $U >$ -3 $k_BT$, and within the considered range, the value of $U$ will not affect our fitting results. $v$ is a local packing parameter *i.e.*, a local volume fraction within the aggregate and for the random packing of polydisperse spheres it does not exceed 0.65 [32–35]. Fig. 2 shows the scattering patterns at 0 and 100 s (Fig. 1D) together with simulated curves based on Eq. 2, in which the silica NPs' contributions were taken from the Monte Carlo fitted form factor (Fig. 1B), whereas the $S_{eff}(q,r_i)$ contribution was included from Eq. 5. The simulations show the important effects that polydispersity has on the structure factor and the position of the correlation peak in ROI V. Typically, for correlations originating from (sticky) hard sphere interactions, one considers the following dependence, for the approximated position of the peak at $q_{max}$:

$$2R_{HS} \approx \frac{2\pi}{q_{max}} \tag{6}$$

Eq. 6 infers that the expected average hard sphere radius, $R_{HS}$, would be equal to the mean radius of a silica particle $<r>$ = ~2.5 nm (Fig. 1B, Table S2 and Eqs. 2&3). This in turn suggests that silica NPs on average touch each other without any LZM molecules in between, or that the protein molecules, if present within the aggregates and among individual silica NPs, are very strongly deformed, likely to a point that they barely contribute to the determined $R_{HS}$. Nevertheless, the simulation in Fig. 2A clearly shows that if the size distribution of silica NPs is actually correctly accounted for, then in order to fit the peak position accurately, an additional "spacer", $aR_{eHS}$ (additional effective hard sphere radius) has to be included in Eqs. 2 & 5:

$$R_{HS,i} = r_i + aR_{eHS} \qquad (7)$$

By setting merely $r_i = R_{HS,i}$ (i.e. $aR_{eHS} = 0$) the position of the simulated peak (from Eq. 5) visibly shifts towards higher-$q$ values with respect to the measured peak. Here, $aR_{eHS}$ is associated with the presence of a single LZM molecule located in between individual silica NPs with the diameter of the LZM molecule represented by $2aR_{eHS}$. The simulation in Fig. 2B also shows that the $v$ packing factor within the aggregates, which directly correlates with the intensity of the broad peak around $q \sim 1$ nm$^{-1}$, has to be relatively high ($v > \sim 0.4$) in order to be able to simulate the intensity profile at $q \sim 1$ nm$^{-1}$ in the later stages ($\sim 100$ s).

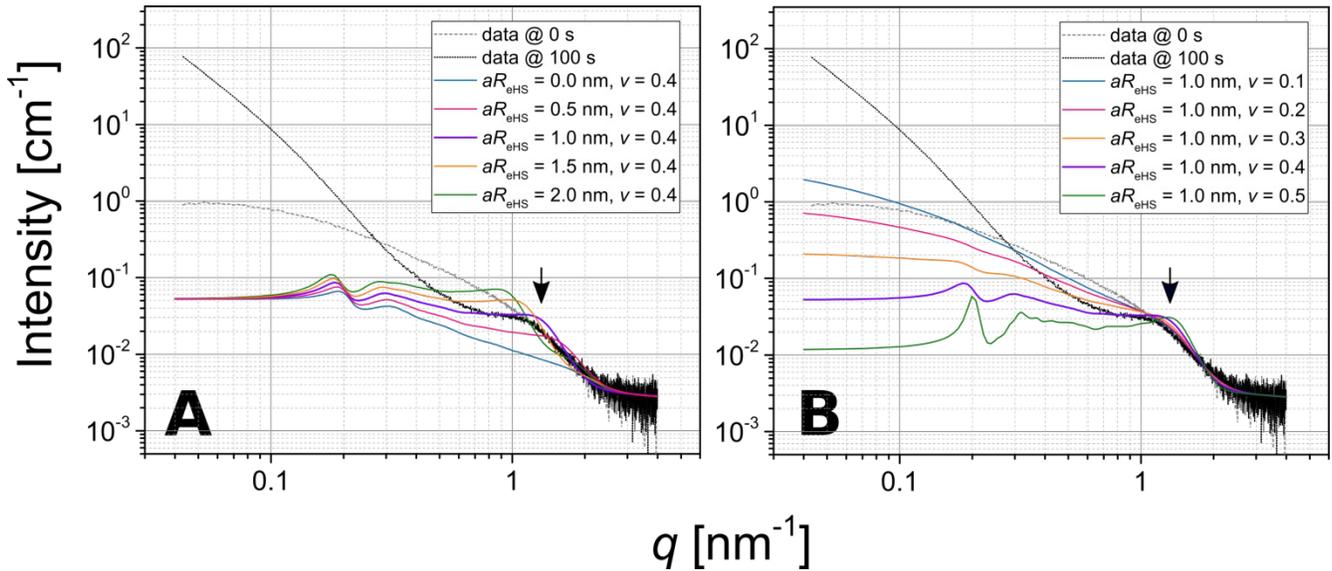

**Figure 2**. Comparison of the measured (black) and simulated (coloured) scattering patterns to illustrate the contribution of the $S_{SHS}(q)$ from Eq. 5, plugged into Eq. 2 as the only structure factor expression. It was the aim to represent the correlation peak at $q \sim 1$ nm$^{-1}$, whereas the effects at low-$q$ are further discussed in the text and in Fig. 3; A) The position of the correlation peak at $q \sim 1$ nm$^{-1}$ (arrow) is reproduced more accurately by $S_{SHS}(q)$ for the polydisperse silica NPs (Fig. 1B), only if one considers an additional effective hard sphere radius, $aR_{eHS}$ (Eq. 7); B) the effect of the local volume fraction $v$ onto the intensity of the correlation peak (arrow). The data uncertainties are not shown for clarity.

*The low-q intensity increase, aggregation, and the structure factor expression of aggregates with internal correlations*

The structure factor from Eq. 5 does not reproduce the observed intensity increase at low-$q$ (Fig. 1 and 2A), because the sticky hard sphere structure factor is derived under the assumption that the interactions extend to an infinite length-scale, with respect to the probed volume. However, in our experiments, we initially have a finite number of "loose" silica NPs that are then rearranged to large aggregates upon mixing with the lysozyme. Hence, microscopically such aggregates must have a finite size, even if their size (radii of gyration) cannot be determined directly from our scattering data, due to the used $q$-range (see Experimental). However, ultimately, to obtain a good fit an expression for $S_{eff}(q)$ (Eq. 2) has to account for both the fine-structure within the aggregates causing the interparticle correlations (as in Eq. 5) and also the low-$q$ intensity increase due to the presence of the interface between the aggregates/composites and the solvent matrix.

A general expression for such an effective structure factor for particles within an aggregate/droplet was proposed originally by Hashimoto et al.[36]. Several variations and applications of this concept are furthermore known e.g. refs[30,37]. For the purpose of the analysis of our own data, we further extended the expressions originating from Hashimoto et al. as we show below in a final form and in the Appendix presenting the complete derivation and the rationale. Our derivation is here essential, because it allowed us to quantify indirectly the changes in the size of the aggregates, although the direct measurement of their radii of gyration was not possible. The so-derived general expression for $S_{eff}(q,r_i)$ (Eq. 8) is expressed as the sum between the structure function of an aggregate ("template"), $S_{agg}(q)$, and the structure factor of the aggregate's internal arrangement, $S_{\text{int}}(q)$, which in our case becomes subsisted by $S_{\text{SHS}}(q)$ (Eq. 5).

$$S_{\text{eff}}(q, r_i) = S_{\text{int}}(q) + S_{\text{agg}}(q) = S_{\text{SHS}}(q, R_{\text{HS,i}}, v, \tau) + A \cdot q^{-D} \tag{8}$$

where $D$ is a fractal dimension describing the arrangement of primary particles within the composites, and $A$ is a single collective fitting parameter in our model, which is proportional to the number density of aggregates, $N_{\text{agg}}$ and their specific surface area, $SSA_{\text{agg}}$. Hence, it expresses indirectly the size/extent of the aggregates. In Fig. 3, we show that the fits with the introduced expressions for the partial structure factor contributions indeed represent the structural features present in the selected scattering pattern (example @ 100 s). However, as is evident from Fig. 3A, the $S_{\text{SHS}}(q)$ from Eq. 5 has to be further improved, because in Fig. 3A the correlation peak is relatively broad ("smeared out"), yet still intense. Typically one would expect such a broadened shape if the local volume fraction parameter, $v$, was smaller than derived from the best fit (i.e., < ~0.4). However, that would also inevitably yield a smaller relative intensity of this peak (see Fig. 2B). Hence, in order to explain this contradiction, one has to remember that the position of the maximum and its shape are predominantly related to $aR_{\text{eHS}}$. The shape of the peak can be modelled substantially better if one allows for a distribution of this parameter in the fitting routine. The need for such mathematical treatment is in fact a manifestation of the actual physical effects, if we consider that $aR_{\text{eHS}}$ represents a radius of a LZM molecule. A LZM molecule can become, at least partially, heterogeneously deformed (on average, in a global sense) within an aggregate e.g., due to the variation of local forces, which in turn is a consequence of polydispersity of the silica NPs, the random character of the packing of the silica NPs etc. More importantly, since lysozyme is a small prolate ellipsoidal protein, with its principal semi-axes being 1.5 nm x 1.5 nm x 2.25 nm the polydispersity in $aR_{\text{eHS}}$ may account for the fact that the protein molecules can be differently orientated during adsorption to NPs. Yet, so far we tried to represent their contribution through a (hard) spherical model. To overcome this, we used a Gaussian distribution to define the average structure factor <$S_{\text{SHS},i}(q)$> in Eq. 9. The application of this structure factor ultimately leads to smearing of the maximum at a constant $v$ and hence yields significantly improved fits (Fig. 3B).

$$\langle S_{\text{SHS},i}(q, r_\text{i} + \langle aR_\text{eHS}\rangle, \sigma, v, \tau)\rangle = \qquad (9a)$$

$$= \frac{\int_0^{\langle aR_\text{eHS}\rangle + 8\sigma} D_\text{G}(\langle aR_\text{eHS}\rangle, \sigma, R) \cdot S_{\text{SHS},i}(q, r_\text{i} + R, v, \tau)\text{d}R}{\int_0^{\langle aR_\text{eHS}\rangle + 8\sigma} D_\text{G}(\langle aR_\text{eHS}\rangle, \sigma, R)\text{d}R}$$

$$D_\text{G}(\langle aR_\text{eHS}\rangle, \sigma, R) = \frac{1}{\sqrt{2\pi\sigma^2}} \exp\left(-\frac{(R - \langle aR_\text{eHS}\rangle)^2}{2\sigma^2}\right) \qquad (9b)$$

In Eq. 9, the mean of the distribution is <$aR_{\text{eHS}}$>, whereas σ denotes a standard deviation. This is the final expression that was then used to represent $S_{\text{int}}(q)$ in Eq. 8 and to fit all the scattering curves from regions II & III in Fig. 1. The numerical integration was performed for each $i$-th bin of the discrete size distribution characterizing the form factor (Fig. 1B). The complete source code and selected data sets are deposited at GitHub.com (see the link in Experimental).

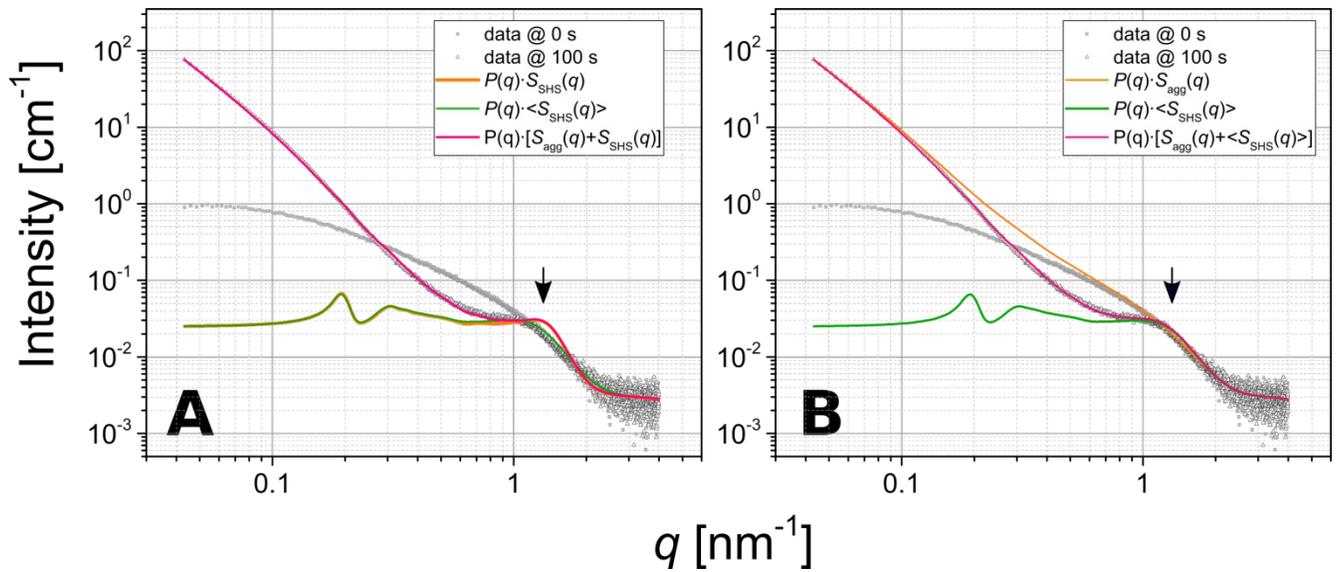

**Figure 3.** Fitting of the structure factor contributions to a scattering curve measured at 100 s during the composite formation processes; A) comparison between the effect of the unsmeared (monodisperse $aR_{eHS}$, red line) $S_{SHS}(q)$ from Eq. 5 and the smeared (polydisperse, green line) $<S_{SHS}(q)>$ from Eq. 9. Fits yielded $v$ = 0.454±0.008, $aR_{eHS}$= 0.979±0.036 nm, $<aR_{eHS}>$= 0.930±0.000 nm, and $\sigma$ = 0.533±0.050 nm; the complete structure factor fit (pink line), which includes the unsmeared $S_{SHS}(q)$ illustrates that the form of the function affects only the correlation peak at $q \sim$ 1 nm$^{-1}$ (arrow), and not the low-$q$ scattering intensity of the aggregates; B) the final fit to the data (pink line) with each structure factor contribution plotted separately (green and orange lines). Parameters for the $<S_{SHS}(q)>$ are the same as in panel A, whereas for the $S_{agg}(q)$ from Eq. 8, we obtained $A$ = 0.0437 ± 0.0001, $D$ = 2.37 ± 0.00. The data uncertainties are not shown for clarity.

*Application of the Scattering Model*

We used the above described scattering model to fit the time-resolved data set and to derive the fitting parameters as a function of time (Fig. 1, Period II, III and beginning of IV, 3010 curves). Fig. 4 shows the time dependence of the five model parameters obtained by fitting the scattering data between 24.5 and 175 s.

During the first ~5 s after mixing, the values of <$aR_{eHS}$> and σ lie way outside reasonable error margins (Fig. 4A & B). This is to be expected, because the corresponding local volume fraction $v$ is very low (<< 0.1, Fig. 4C) at the beginning of the composite formation process (i.e., the beginning of period II in Fig. 1). Consequently, the contribution of <$S_{SHS}(q)$> to the structure factor during this period is mostly negligible with respect to $S_{agg}(q)$. As $v$ reaches ~0.1 at ~30 s (transition between regions II and III in Fig. 1) the actual evolution of <$aR_{eHS}$> and σ begin: <$aR_{eHS}$> starts at ~0.7 nm and rapidly increases to ~0.9 nm by ~50 s and then more gradually to ~ 1.1 nm by 150 s (period III Fig. 1). The associated standard deviation σ follows a similar trend as <$aR_{eHS}$>, growing from 0.35 nm to 0.5 nm by 50 s, and then levels off at ~0.5 nm by 150 s (end of period III in Fig. 1A) within the fitting uncertainty. These initial rapid changes up to 50 s are also reflected in the evolution of parameter $v$ (Fig, 4C), where the parameter rapidly increases to ~0.45 and then remains constant within the fitting uncertainty up to 150 s. Because <$aR_{eHS}$> directly relates to the size of lysozyme within the silica aggregate, we can link the changes in <$aR_{eHS}$> to possible changes in the protein's structure/folding/shape. Between ~30 and ~50 s, where the LZM molecules rapidly induce the aggregation of the silica NPs, the LZM molecules appear to undergo a deformation (compression). Yet, as this process approaches equilibrium, through the internal densification of the aggregates, the molecules gradually return to their native dimensions. This result is in agreement with findings concerning the activity of lysozyme within composites with silica, showing that smaller silica NPs (as those used for our experiments) promote higher enzymatic activity of lysozyme, and that this in turn is dependent on the preservation of the native shape of the molecule upon composite formation[11].

Furthermore, the formation of the silica-LZM composite is dominated, from the very moment of mixing for the initial 20 s, by a rapid, 4.5 fold increase of parameter $A$ (Fig. 4D). This is best explained by the increasing number density of the aggregates, $N_{agg}$, and the associated increasing specific surface area, $SSA_{agg}$ (see Eq. 8, and Eq. 19 in the Appendix). After $t = 40$ s, parameter $A$ further increased, albeit at a slower rate. The concurrent evolution of the fractal dimension (parameter $D$; Fig. 4E) suggests that initially (up to 50 s), the aggregates have a relatively open morphology with $D$ values < 2.4 and characterized by a limited contribution of $<S_{SHS}(q)>$ due to $v$ < 0.1 (Fig. 4C). Afterwards (> ~50 s), the aggregates reached an internally denser state, as reflected by the steadying of both values for $D$ (~2.4) and $v$ (~0.45). In other words, since these two parameters, $D$ and $v$, reflect the internal structure of the aggregates from the perspective of the two structure factor contributions (Eq. 8 above and Eq. 19 in the Appendix), their evolution clearly indicates no further internal changes in the aggregates between ~50 and 150 s. If such an internal densification processes had occurred, one would expect that it would have contributed to the decrease of the specific surface area of the aggregates, $SSA_{agg}$. Interestingly however, parameter $A$ (Fig. 4D), keeps increasing after 50 s, i.e., after the internal dense structure is established, meaning that the product of number density of the aggregates and their specific surface area actually increased. This is possibly a result of a secondary processes involving the "breakup" of larger aggregates into smaller units. Indeed, if we correlate the changes in $A$ and $v$ (Fig. 4F), we observe three stages of such a secondary processes. In the first stage for $v$ < 0.1 (up to 25.5 s), $A$ grows as a function of $v$ in a bound exponential mode, which translates into an increasing number of low-dimensional aggregates with hardly any internal correlations, forming an extended network of particles of low dimensionality $D$ (Fig. 4E). In the second stage as $v$ increases from 0.1 to ~0.45 (25.5 s to ~70 s), $A$ vs. $v$ (Fig. 4F) shows a linear dependence, indicating that as the number density of aggregates increases, they also gradually densify, and that the growth of the aggregates occurs at the same rate as their internal densification. Finally in the third stage, once $v$ remains relatively constant at ~0.45 (after 70 s), the product of the number density and the specific surface area of the aggregates continues to increase as documented by the increasing *A*, yet without any further dramatic changes to the internal structure/arrangement (i.e. constant *D* and *v*), implying the aforementioned breakup of the larger aggregates into smaller units. In overall, this processes can be best explained as the initial rapid flocculation/clumping of NPs and LZM together into an extensive network just after mixing as the system is out of equilibrium, followed by the gradual evolution towards a steady state, in which smaller aggregates are more favourable.

During period IV (> 150 s), the time evolution of the three parameters (<$aR_{eHS}$>, $v$ and $\sigma$; Fig. 4A-C) exhibited a characteristic discontinuity from the trends observed during periods II and III. This is because at time > 150 s, the scattering intensity at high-$q$ (which corresponded to the form factor, Fig. 1D) changed significantly, so that the original form factor of pre-mixing silica NPs from Fig. 1B was not representative for silica particles after 150 s. Thus, we could no longer use the fitted size distribution in our model, and any trends of these three parameters (Fig. 4A-C) were not valid any more after 150 s. On the other hand due to the fact that the low-$q$ part of the data by definition is practically independent from the form factor, in fact the evolution of parameters *A* and *D* (trends in Fig. 4D&E), even after 150 s are representative for the processes at the length scales corresponding to entire aggregates. However, due to the fact that in period IV our scattering is no longer self-consistent, we do not analyse those trends.

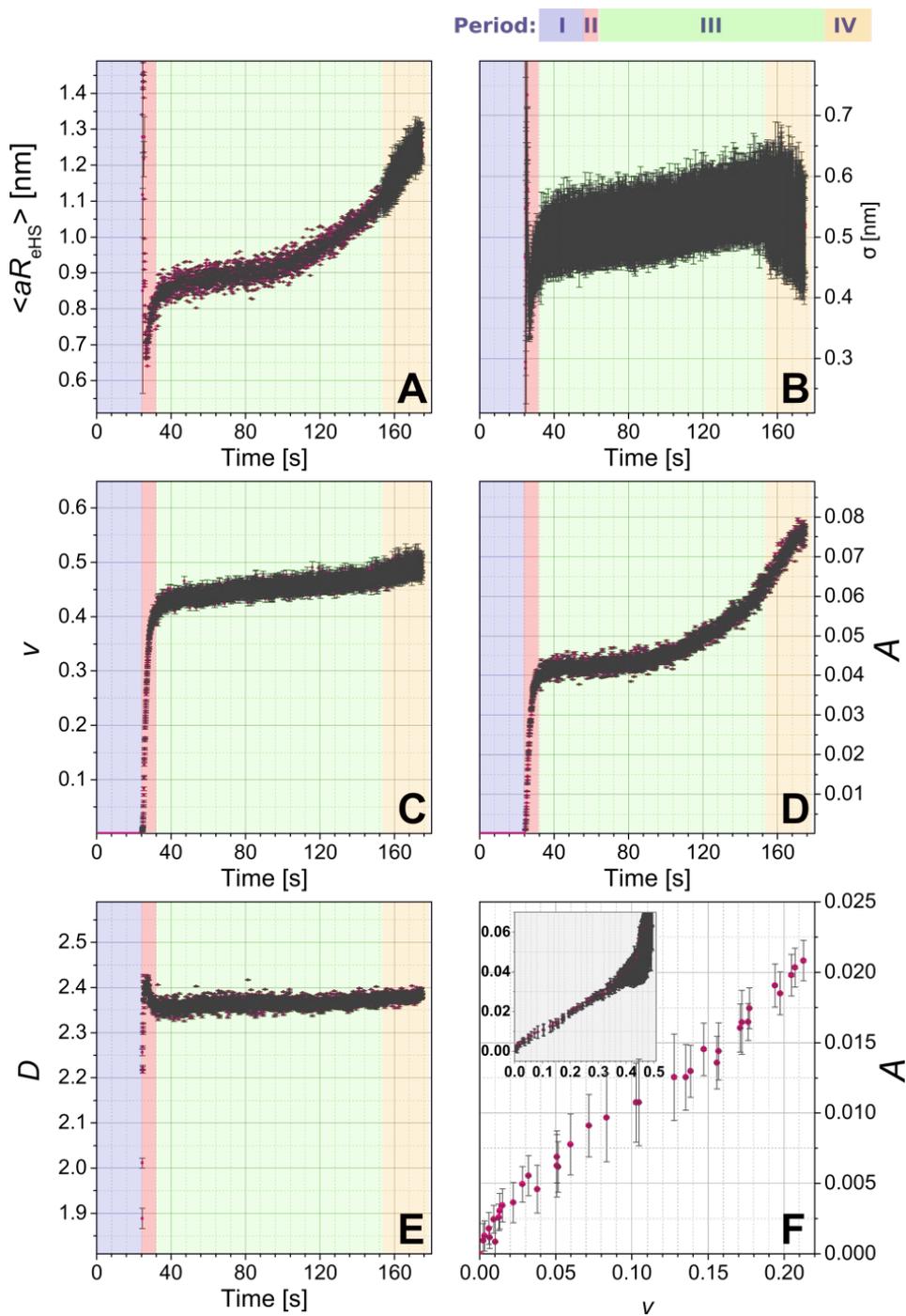

**Figure 4.** Evolution in the crucial parameters of the fitting model plotted as a function of time; A) mean additional hard sphere radius <$aR_{eHS}$>; B) the corresponding standard deviation σ; C) local volume fraction $v$; Parameters characterising <$S_{agg}(q)$>: D) $A$, relative product of the number density of the aggregates $N_{agg}$ and their specific surface area $SAA_{agg}$; E) dimensionality of the aggregate $D$; F) correlation $A$ vs. $v$, with inset showing the same as the main figure 4F, but for a wider range; axes legends in the inset correspond to those in the main figure.

**Implications**

The analysis of the evolution of the fitting parameters (Fig. 4) paints a clear image of the four-step sequence of events during an aggregation induced by the interaction between the protein LZM and amorphous silica NPs (Fig. 5). Immediately upon mixing, aggregation is induced due to the opposing surface charge of the silica NPs and the protein (Fig. 5A). An infinitely extensive and open ($D$=~1.8-2.2) aggregate network, from the point of view of the SAXS measurement, forms within ~4 data frames (~200 ms). The so-formed network initially has no internal correlations, as is expected for a classical mass fractal[38]. However soon after (~1 s), areas of correlated NPs-LZM domains start forming within the network and the increase in the internal volume fraction, $v$, indicates an internal densification and ordering (Fig. 5B). This is also reflected by the fact that parameter $D$ reaches a stable and relatively high value of ~2.4, which is characteristic for denser mass fractals. Such a fractal dimension for silica-lysozyme aggregates was previously reported[5,12] and can be associated with the diffusion limited particle-cluster aggregation (DLPCA) mechanism[39-44]. The DLPCA growth mode is also evidenced not only by the plateau value of $D$, but also through the fact that the value of $D$ increases with the size/extent of the aggregates[39,41] expressed indirectly by the parameter $A$ (see region II in Figs. 4D&E). This means that aggregates grow through the accretion of individual primary particles to larger aggregates[44], where aggregates as such become denser as their size increases, which in consequence favours the eventual occurrence of the correlated domains (which are the ultimate dense regions in the aggregate composed of smaller particles).

The parameters characterizing the interparticle correlation effects carry information about the size of the LZM molecules bridging our silica NPs. From Fig. 4A&B it is clear that in this network the dimensions of the protein molecules are considerably smaller than the native dimensions of lysozyme in any possible orientation. Hence, this suggests that initially the binding of silica NPs by lysozyme involves a severe deformation/unfolding of the protein molecules, followed in time by a relaxation and increase in the protein dimensions towards a (more) native state (Fig. 5C). The final value of the radius of 1.1±0.5 nm for the protein, which is reached before 150 s, is close to a radius of the protein in a side-on orientation (~1.5 nm), rather than in the end-on orientation (~2.25 nm). Our *ex situ* analysis of the dried composite samples (see Experimental) showed that at 1000 ppm lysozyme, 32.7% wt. of the protein was incorporated into the composites. This means that for silica NPs precipitated at the concentration of 0.8737 g/L (calculated from the volume distribution in the SAXS patterns), the concentration of lysozyme in the composite was 0.4245 g/L. This is valid under the assumption that all available silica NPs were bound in aggregates with lysozyme. Hence, the number density of the protein molecules was $N_{LZM}$ = ~1.8e19 L$^{-1}$, and for silica $N_{NP}$ = ~1.8e19 L$^{-1}$ (from SAXS by converting the volume distribution to a number distribution). This directly suggests that the silica-lysozyme aggregates are near-stoichiometric, with 1 protein molecule associated with 1 silica NP. Such a stoichiometric relationship is actually expected for small silica particles of the size close to the one of the protein molecule.[11,45] Su et al.[46] found that at small surface coverage the lysozyme attaches to silica NPs in a side on orientation, and recently the molecular dynamics simulations by Hildebrand et al.[47] also further confirmed that the side-on orientation of lysozyme with respect to silica constitutes the configuration of the highest attraction. This together with the relatively low dipole moment and the positive surface charge of the protein surface, potentially account for the bridging of the NPs by the LZM molecule, as the protein does not show a favoured orientation of the opposite active sites in the side-on orientation (i.e. both active sites show similar binding properties). In such a case one should indeed expect the DLPCA mode of aggregation, with the binding of the protein to the silica NP surfaces taking place through specific amino acids at the opposite sides of the molecule[13,47–49]. The densification of the internal structure of the aggregates reaches a steady point, when the LZM molecules relax to their native-like dimensions. Yet, at the same time the actual network constituting the composite, appears to break up into smaller aggregate units. The morphological changes of the composites further continue beyond 150 s. This is documented through the change in the form factor of the silica NPs which appear to grow in size, compared to the pre-mixed initial NPs. Although we cannot use our model to explain this last stage, we can speculate that the observed change is caused by a partial coalescence of NPs inside of the aggregates.

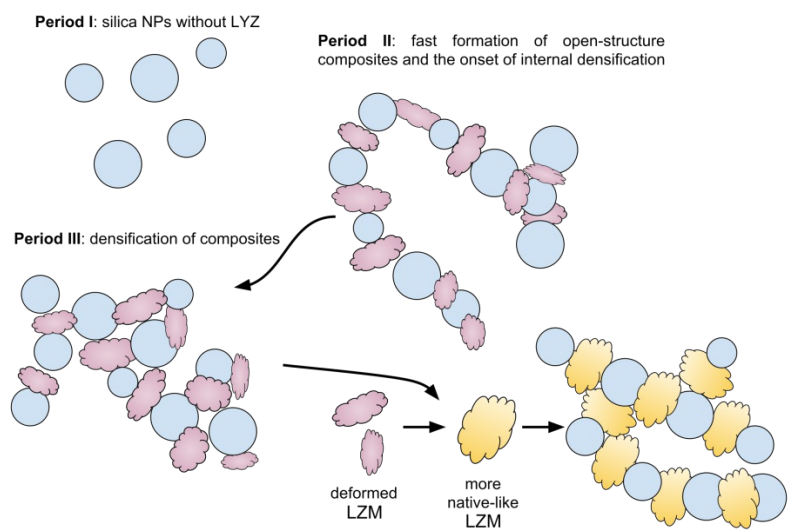

**Figure 5.** Schematic of the selected steps in the formation of silica-LZM composites as derived from the fast *in situ* and time resolved SAXS data. Period I - unaggregated silica NPs (spherical in shape blue objects) before LZM addition. Following the LZM addition (cloud-like purple and yellow objects); Period II - fast aggregation to low-dimensionality fractal network constituting the internal structure of large composite particles (primarily $S_{agg}(q)$ contribution); Period III - internal densification of the network in the process of which the interparticle correlations develop (evolution of $<S_{SHS}(q)>$ and $S_{agg}(q)$). At the early stages of densification the lysozyme molecules are strongly deformed within the aggregates; further densification during which the protein molecules appear to relax to their more-native dimensions.

**Conclusions**

The analysis of *in situ* scattering data collected at 20 fps from the formation of silica-lysozyme composites showed that the processes involved formation of large aggregated structures in which individual silica NPs were bridged by LZM molecules. We developed and applied a new scattering model to underpin the changes of the morphology of the composites as a function of time. This model allowed us to unravel that the formation follows diffusion limited particle-cluster aggregation (DLPCA) mechanism, which results in relatively densely-packed mass-fractal-like aggregates within which non-fractal correlated domains of particles evolve. Furthermore, we used the scattering model to link the evolution in the measured structure factor to the lysozyme molecule, and we found that the aggregation processes involve severe deformation of the protein molecules, which is then followed by the relaxation towards the original dimensions.

**Appendix**

*1. Derivation of Eq. 8*

The contribution of $S_{\text{int}}(q)$ in Eq. 8 has a rather simple form, however the actual meaning of the *A* parameter extends beyond this short representation. Below, we present the derivation of the equation and we indicate the approximations we make on the way to the final form of the equation.

First we consider a structure factor function describing an arrangement of primary particles of a certain scattering length density (SLD), $\rho_p$ within an aggregate. The SLD of the space in between the particles within the aggregate (the "template") is $\rho_t$, whereas the SLD of the surrounding ("solvent") is $\rho_s$. The average scattering contrast of the aggregate, $\Delta\rho_{\text{agg}}$ is:

$$\Delta\rho_{\mathrm{agg}} = v\left(\rho_{\mathrm{p}} - \rho_{\mathrm{s}}\right) + (1-v)\left(\rho_{\mathrm{t}} - \rho_{\mathrm{s}}\right) \tag{10}$$

where *v* is a local volume fraction of particles within the aggregate (as in Eq. 5). We also define any contrast fluctuation within an aggregate as:

$$\Delta\eta = \rho_{\mathrm{p}} - \rho_{\mathrm{t}} \tag{11}$$

If we write the structure function of an aggregate ("template") as $S_{\mathrm{agg}}(q)$ and the internal arrangement within this aggregate as $S_{\mathrm{int}}(q)$, then similar to Hashimoto et al.[36] and Lin et al.[37] we can express the scattering intensity by the generalized Eq. 12, in which $\otimes$ denotes a convolution operation of the functions.

$$S_{\mathrm{eff}}(q) = S_{\mathrm{agg}}(q) + \frac{V_{\mathrm{p}} \cdot \phi_{\mathrm{p}} \cdot \Delta\eta^2}{V_{\mathrm{agg}} \cdot \phi_{\mathrm{agg}} \cdot \Delta\rho_{\mathrm{agg}}^2} S_{\mathrm{agg}}(q) \otimes S_{\mathrm{int}}(q) \approx S_{\mathrm{agg}}(q) + \frac{V_{\mathrm{p}} \cdot \phi_{\mathrm{p}} \cdot \Delta\eta^2}{V_{\mathrm{agg}} \cdot \phi_{\mathrm{agg}} \cdot \Delta\rho_{\mathrm{agg}}^2} S_{\mathrm{int}}(q) \tag{12}$$

where $V_{\mathrm{agg}}$ and $\phi_{\mathrm{agg}}$ represent a volume and a volume fraction of aggregates in the solution, respectively. Here, $S_{\mathrm{int}}(q) = S_{\mathrm{SHS}}(q)$ from Eq. 5, whereas the $S_{\mathrm{agg}}(q)$ contributes to the increase in intensity at low-*q* in the course of the formation of aggregates. The approximation in Eq. 12 is valid when the overall radii of the aggregates ($\xi$) are significantly larger that the interparticle correlation distance from Eq. 5 (i.e. $\xi \gg R_{\mathrm{HS}}$). Furthermore, we must re-normalize $S_{\mathrm{eff}}(q)$ in a different way than Hashimoto et al. and Lin et al. in order to combine it with the form factor defined in Eq. 1 as for our scattering data the intensity for $q \to 0$ in a Porod representation does not level off to a finite value (e.g. Figs. 2&3). Consequently, it is impossible to determine $V_{\mathrm{agg}}$. On the other hand the primary particles' radii and consequently their volume, $V_{\mathrm{p}}$, are known, and therefore it is reasonable to normalize $S_{\mathrm{eff}}(q)$ with respect to the primary particle volume rather than the aggregate volume:

$$S_{\mathrm{eff}}(q) = S_{\mathrm{int}}(q) + \frac{V_{\mathrm{agg}} \cdot \phi_{\mathrm{agg}} \cdot \Delta\rho_{\mathrm{agg}}^2}{V_{\mathrm{p}} \cdot \phi_{\mathrm{p}} \cdot \Delta\eta^2} S_{\mathrm{agg}}(q) \tag{13}$$

Now let us assume that the aggregates follow mass-fractal behaviour and that the correlation function of mass-fractals is described[50,51] by Eq. 14.

$$g_{\text{agg}}(r) = K \cdot r^{D-d} \qquad (14)$$

where $D$ is a fractal dimension, $d$ is the Euclidean dimension ($d = 3$) and $K$ is a normalization constant that is proportional to the mass and surface area of an aggregate. We do not include a cut-off function since in our case the intensity did not level off at low-$q$ as mentioned above. For the sake of simplicity let us assume again that the electron density of the aggregates is homogeneously distributed and its corresponding correlation function only involves a two-phase system. In this case the specific surface area of aggregates ($SSA_{\text{agg}}$) is proportional to the derivative of the correlation function at the near-zero length-scale $r \to 0$:

$$SSA_{\text{agg}} = -4 \cdot \phi_{\text{agg}}(1-\phi_{\text{agg}}) \cdot \lim_{r \to 0} \frac{\text{d}g_{\text{agg}}(r)}{\text{d}r} = (3-D) \cdot \phi_{\text{agg}}(1-\phi_{\text{agg}}) \cdot K \cdot \lim_{r \to 0} r^{D-4} \qquad (15)$$

This is valid for the considered small volume fractions (1- $\phi_{\text{agg}}$) ≈ 1. Since $D < 4$, $SSA_{\text{agg}}$ becomes increasingly larger for decreasing length-scales, yielding an infinite surface area at an infinitely small length-scale $r$. However, since the aggregates are composed of primary particles with a typical radius $R_{\text{HS}}$, we can say that the aggregate does not contain smaller features than those primary particles (i.e. $r \geq R_{\text{HS}}$). Therefore, we find a finite specific surface area for mass fractal aggregates:

$$SSA_{\text{agg}} = (3-D) \cdot \phi_{\text{agg}} \cdot K \cdot \lim_{r \to 0} r^{D-4} = (3-D) \cdot \phi_{\text{agg}} \cdot K \cdot R_{\text{HS}}^{D-4} \Rightarrow$$

$$\Rightarrow K = \frac{\sigma_{\text{agg}} \cdot R_{\text{HS}}^{4-D}}{(3-D) \cdot \phi_{\text{agg}}} \qquad (16)$$

and this way this newly derived $K$ constant in Eq. 16 substitutes the $K$ constant from Eq. 14 and thus, we can use the correlation function from Eq. 14 to calculate the structure factor. Please note the Hashimoto et al. described the structure factor in such way that it is normalized as a form factor, i.e., it is normalized by its total volume. This is in line with the structure factor of mass fractal aggregates as described by Sorensen and Wang[50], yet it is different from a better-known derivation by Teixeira[52]. Both Sorensen and Wang, and Teixeira's approaches are valid as long as one considers normalizations explicitly. The structure factor is described by the rotation-averaged Fourier transform:

$$S_{\text{agg}}(q) = \frac{N_{\text{agg}}}{V_{\text{agg}}} \int_0^\infty g_{\text{agg}}(r) \cdot 4\pi r^2 \frac{\sin(qr)}{qr} \, \text{d}r =$$

$$= \frac{4\pi \cdot \Gamma(D-1) \cdot N_{\text{agg}} \cdot SSA_{\text{agg}} \cdot R_{\text{HS}}^{4-D}}{V_{\text{agg}} \cdot \phi_{\text{agg}} \cdot (3-D)} \cdot \sin\left(\frac{(D-1)\pi}{2}\right) \cdot q^{-D} \quad (17)$$

where $N_{\text{agg}}$ is the number density of aggregates and $SSA_{\text{agg}}$ is their specific surface area. By substitution of Eq. 17 into Eq. 13, we obtain:

$$S_{\text{eff}}(q) = S_{\text{int}}(q) + \frac{\Delta\rho_{\text{agg}}^2}{\Delta\eta^2} \cdot \frac{4\pi \cdot \Gamma(D-1)}{(3-D)} \cdot \frac{N_{\text{agg}} \cdot SSA_{\text{agg}} \cdot R_{\text{HS}}^{4-D}}{V_{\text{p}} \cdot \phi_{\text{p}}} \cdot \sin\left(\frac{(D-1)\pi}{2}\right) \cdot q^{-D} \quad (18)$$

For spherical primary particles, $V_{\text{p}}$ is known and hence this way our derivation leads us to a final form of the equation for the effective structure factor:

$$S_{\text{eff}}(q) = S_{\text{int}}(q) + \frac{\Delta\rho_{\text{agg}}^2}{\Delta\eta^2} \cdot \frac{\Gamma(D-1)}{(3-D)} \cdot \frac{N_{\text{agg}} \cdot SSA_{\text{agg}}}{\phi_{\text{p}} \cdot R_{\text{HS}}^{D-1}} \cdot \sin\left(\frac{(D-1)\pi}{2}\right) \cdot q^{-D} =$$

$$= S_{\text{int}}(q) + A \cdot q^{-D} \quad (19)$$

In Eq. 19 we introduced several simplifications. Firstly, $N_{\text{agg}}$, $SSA_{\text{agg}}$, $\Delta\eta$ and $\Delta\rho_{\text{agg}}$, and are essentially unknown, and it is impossible to determine any one of these parameters independently. They have to be combined into a collective parameter. This is necessary as Eq. 19 was derived for a system characterized by a monodisperse particle distribution with only a single value of $\phi_p$ for a given $R_{HS}$ and the resulting $N_{agg}$. For a polydisperse case, as is the case in our study, although a population of primary particles is described by $D(r_i)$ (Eq. 2), the resulting distribution of sizes of aggregates will be totally independent from this initial distribution, and it will also be unpredictable. Secondly, the $R_{HS}^{1-D}$ component in Eq. 19 could be potentially important, since it determines the high-$q$ cut-off at which the contribution of the structure to the intensity lessens, and the form factor dominates. However, in Eq. 19 this very transition point is dominated by a $S_{int}(q)$ contribution and its strong correlation peak. Hence, we assume that $R_{HS}^{1-D} \sim 1$. Thirdly, the remaining part of the expression depending on parameter $D$ is practically constant at a value ~1.2, and although we could introduce it explicitly in the model it does not affect the final trends. Hence, as a result of the above approximations, we use $A$ as a single collective fitting parameter in our model. Changes is $A$ therefore should be interpreted primarily as the average change of the product of the number density of aggregates and their specific surface area, and through these two physical parameters are related to the size (or "extent") of the aggregates.

## 2. Supporting tables and figures

**Table A1.** Selected physicochemical properties of the initial amorphous silica used in the experiments

| | |
|---|---|
| **Molecular mass [g/mol]** | 60.08 |
| **Density [g/cm$^3$]** | 2.196 |
| **Scattering length density, SLD [Å$^{-2}$]** | $1.883 \cdot 10^{-5}$ |
| **ΔSLD with respect to H$_2$O [Å$^{-2}$]** | $9.362 \cdot 10^{-6}$ |
| **Solid amount* from a 1000 ppm a sodium metasilicate solution at pH=7.5 and $T$ = 21 deg. C [mmol/L]** | 14.85 mmol/L, equivalent to 0.041% volume fraction |

*calculation performed in PHREEQC [25]

**Table A2.** Silica size distribution from fitting

| | |
|---|---|
| **Mean radius [nm]** | 2.525±0.011 |
| **Variance [nm$^2$]** | 3.707±0.029 |
| **Skew** | 3.976±0.081 |
| **Kurtosis** | 24.80±1.21 |

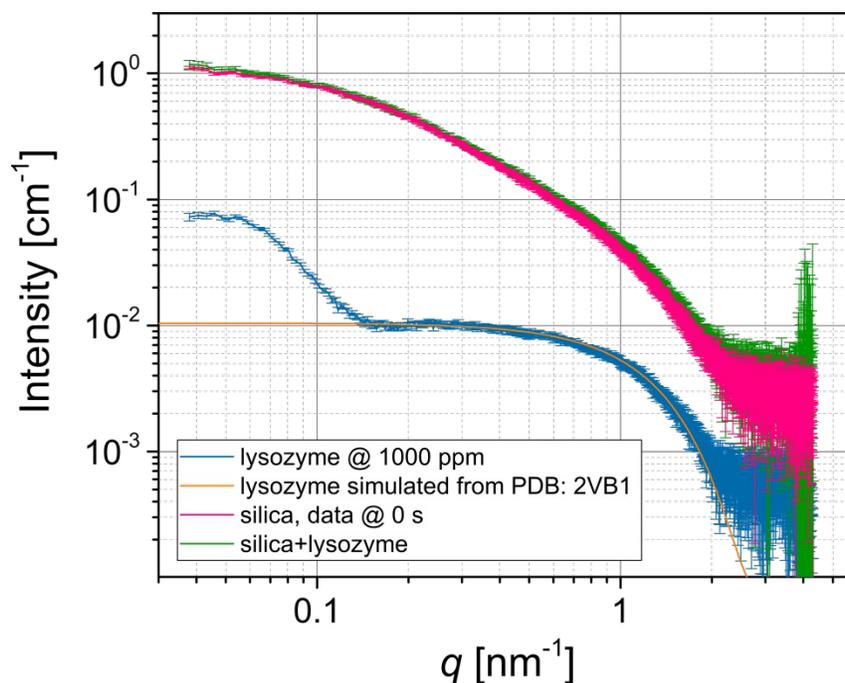

**Figure A1.** Scattering curves of solutions with the initial amorphous silica NPs and the LZM solutions measured independently prior to mixing. The pattern for lysozyme at 1000 ppm (the blue curve) matches well the simulated scattering pattern (the orange curve) generated from the PDB structure file[53] 2VB1, apart from at $q < 0.15$ nm$^{-1}$. The observed low-$q$ intensity increase in the LZM solution originates from a very small population of larger particles/clusters/aggregates and is negligible. This can be demonstrated by simply adding together the scattering intensities from the initial silica NP solution @ 0 s (the pink curve) and the LZM solution (the blue curve) and accounting for the uncertainties. The resulting pattern (the green curve) is within the uncertainty indistinguishable from the scattering of the silica NPs solution on its own. Such addition corresponds to the hypothetical case when there were no interactions between the silica NPs and the protein, but it also points out that despite relatively high protein concentration (with respect to silica) the resulting scattering intensity of the protein is very low (see also ref. [16]).

**Conflicts of interest**

There are no conflicts of interest to declare.

**Acknowledgments**

This research was made possible by two Marie Curie grants from the European Commission: the NanoSiAl Individual Fellowship, Project No. 703015 and the MINSC Initial Training Research network, Project No. 290040. We also acknowledge the financial support of the Helmholtz Recruiting Initiative grant No. I-044-16-01 and we thank EMBL for granting us beamtime at BioSAXS beamline P12 of PETRA III.